\documentclass[aps]{revtex4}
\usepackage{amsmath}
\usepackage{amsfonts}
\usepackage{graphicx}
\usepackage{longtable}
\usepackage{multirow}
\begin{document}
\title{Nonlinear spectroscopy with entangled photons; manipulating quantum pathways of matter.}
\author{Oleksiy Roslyak}
\email{oroslyak@uci.edu}
\affiliation{Chemistry Department, University of California, Irvine, California 92697-2025, USA}
\author{Christoph A. Marx}
\email{cmarx@ucie.edu}
\affiliation{Chemistry Department, University of California, Irvine, California 92697-2025, USA}
\author{Shaul Mukamel}
\email{smukamel@uci.edu}
\affiliation{Chemistry Department, University of California, Irvine, California 92697-2025, USA}
\date{\today}
\begin{abstract}
Optical signals obtained by the material response to classical laser fields are given by nonlinear response functions which can be expressed by sums over various quantum pathways of matter. We show that some pathways can be selected by using nonclassical fields, through the entanglement of photon and material pathways, which results in a different-power law dependence on the incoming field intensity. Spectrally overlapping stimulated Raman scattering (SRS) and two-photon-absorption (TPA) pathways in a pump probe experiment are separated by controlling the degree of  entanglement of pairs of incoming photons. Pathway-selectivity opens up new avenues for mapping photon into material entanglement. New material information, otherwise erased by interferences among pathways, is revealed.
\end{abstract}
\maketitle
\par
Entanglement is one of the most fascinating manifestations of quantum mechanics. Entangled particles act as a single object even when they are spatially well separated, thus forming a "Schr\"{o}dinger cat". Measurements with entangled particles have been used to test for the foundations of quantum mechanics by resolving the Einstein-Podolsky-Rosen (EPR) paradox and demonstrating Bell's inequalities \cite{haroche2006eqa,PhysRevLett.47.460}.
Many applications to quantum information processing, secure communication (teleportation) \cite{kimble2004qpp}, and quantum computing are based on
the manipulation of entangled particles \cite{PhysRevLett.81.3631,walther2004bwn}. Photon entanglement is easier to create and maintain by e.g. parametric down conversion \cite{mandel1995oca} or bi-exciton decay \cite{PhysRevLett.84.2513,PhysRevLett.39.691} than that of material
particles\cite{edamatsu2007epg}. Most optical applications of entanglement tune the photon frequencies away from material resonances. Nonlinear coupling between optical modes is then described by an effective optical field interaction Hamiltonian. The matter plays only a passive role by
providing the interaction parameters through its susceptibilities but its degrees of freedom do not actively participate in the process.
A clear signature of entanglement in nonlinear optics is that the near-resonant sum frequency generation signal scales linearly (rather than quadratically) with the incoming field intensity. This effect which indicates that the two photons effectively act as a single particle \cite{javanainen1990lid,PhysRevA.31.2409,dayan2007ttp} has been predicted and verified experimentally by several groups \cite{PhysRevB.69.165317,teich1998epm,saleh1998epv,pe'er:073601,dayan:043602,lee2006epa,lee2007qso}. Improved interferometric resolution by entanglement has been demonstrated \cite{agarwal}.
\par
Here we show that nonlinear resonant interactions between entangled photons and matter can lead to a much more dramatic effect; quantum pathway selection. In resonant processes the matter actively participates and gets entangled with the photons, making it possible to control the pathway of matter by manipulating the photons. Pathway interference may be controlled by varying the degree of entanglement, thereby improving the resolution of nonlinear spectroscopic techniques.
\par
Nonlinear optical signals induced by classical optical fields are given by sums of terms which represent different possible quantum pathways \cite{mukamel1995pno}. The signal is related to the expectation value of the dipole operator $\langle \psi(t) | V | \psi(t) \rangle$. The pathways are created by allowing various optical pulses to interact with the ket $|\psi(t) \rangle$ and the bra $\langle \psi(t) |$ in all possible orders. Each technique has its characteristic set of paths which are determined by the wavevectors, carrier-frequencies and bandwidths of the various incoming pulses. These pathways can have different signs and overlap spectrally. Quantum interference among these pathways often results in the elimination of some resonances or the creation of new resonances. This complicates the interpretation of the signals. Non-classical optical fields such as entangled photon pairs, may be used to control and select these pathways. We consider resonant nonlinear processes where matter paths get entangled with photon paths (i.e. each material pathway may be accompanied by a different optical pathway \cite{marx2008nos}), paving the way for active matter-pathway control and selection by manipulating the photons. Entangled photons can be designed to select a subset of pathways, providing new ways for controlling nonlinear optical signals and  interference effects.
\par
\begin{figure}
  \includegraphics[width=10cm]{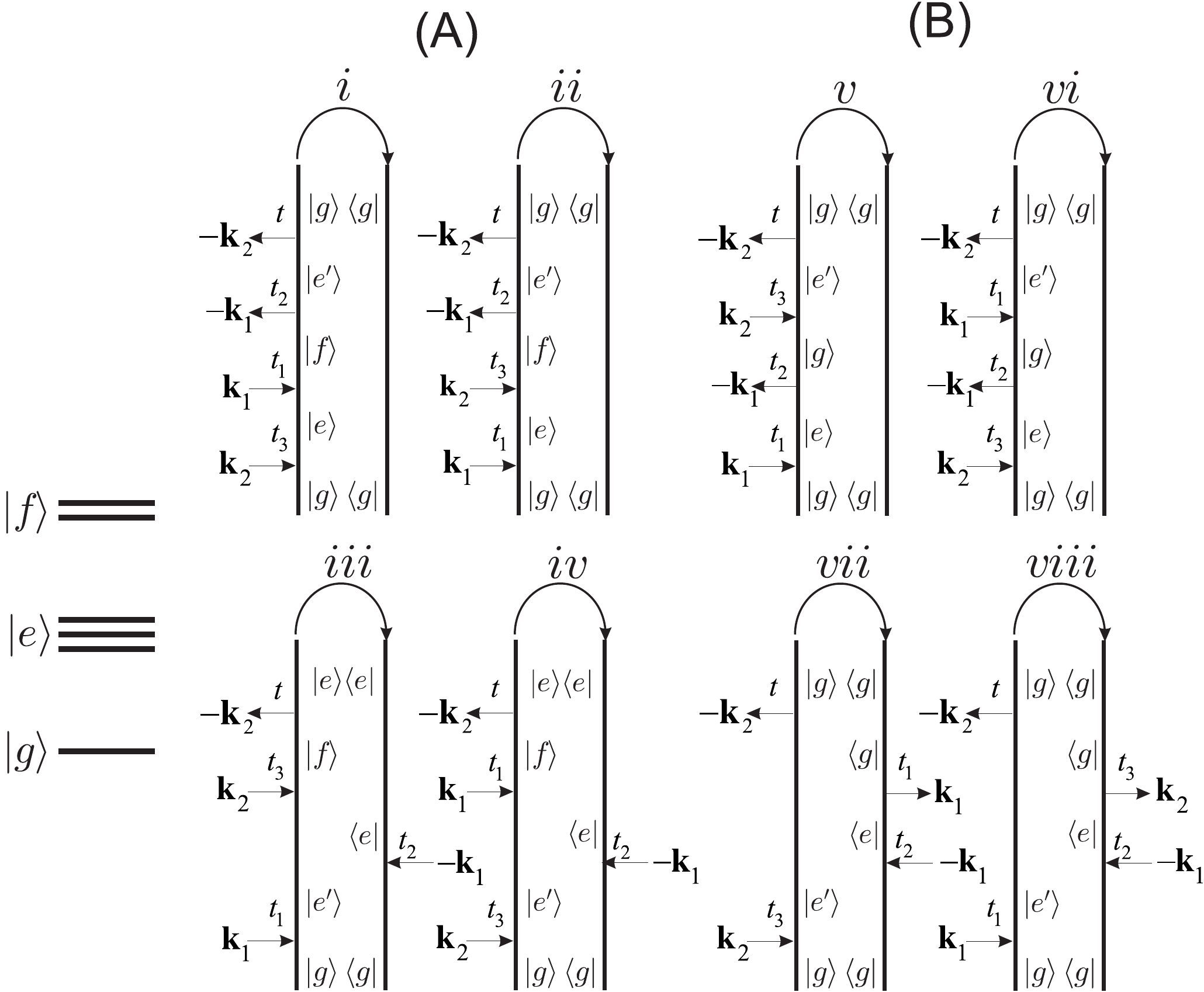}\\
  \caption{Loop diagrams representing the entangled photon-matter quantum pathways contributing to the pump-probe signal. Moving clockwise along the loop we see the sequence of photon absorption (emission) events and the corresponding state of matter.  Pathways (i)-(iv) constitute Group A which contains TPA resonances (absorption, absorption, emission, emission), whereas (v)-(viii) are group B which contains SRS resonances (absorption, emission, absorption, emission).}
\label{FIG:1}
\end{figure}

The pump-probe technique is carried out with two optical modes with wave vectors $\mathbf{k}_1,\mathbf{k}_2$ and frequencies $\omega_1, \omega_2$ interacting with an assembly of $N$ three level $(|g \rangle, | e \rangle , | f \rangle )$ molecules (Fig.\ref{FIG:1}). The levels form a ladder and only $|g \rangle$ to $| e \rangle$ and $|e \rangle $ to $ | f \rangle$ optical transitions are allowed. The signal, defined as the change
in the absorption of $\omega_2$ due to the interaction with $\omega_1$, is given by the sum of the eight quantum pathways depicted by the loop diagrams shown in Fig.\ref{FIG:1}. The left branch represents the ket $|\psi(t) \rangle$ and the right branch stands for the bra $\langle \psi(t) |$. The light/matter interactions occur sequentially starting at the bottom left and moving along the loop clockwise.
\par
\par
We shall divide the pathways into two groups. In group A there are two photon-absorption events followed by two emissions. In these pathways the system evolution along the loop goes as $g \to e \to f \to e \to g$. We shall, therefore, label them two-photon-absorption \cite{glauber2007qto,javanainen1990lid,PhysRevA.31.2409} (TPA) and denote their contribution as $ \chi^{(3)}_{A}$. All of these terms are multiplied by the field correlation functions of the form $ \langle a^\dag a^\dag a a \rangle$, where $a(a^\dag)$ is the annihilation (creation) of the optical modes. As can be seen from the diagrams, the optical field factors for pathways (i-iv) are $\langle{a^\dag_2 a^\dag_1 a_1 a_2}\rangle$, $\langle{a^\dag_2 a^\dag_1 a_2 a_1}\rangle$, $\langle{a^\dag_1 a^\dag_2 a_2 a_1}\rangle$, $\langle{a^\dag_1 a^\dag_2 a_1 a_2}\rangle$ respectively.
\par
In group B the system goes back and forth between single photon absorption and emission, and the material path is $g \to e \to g \to e \to g$. The contribution of these stimulated Raman scattering pathways \cite{mukamel1995pno} to the signal will be denoted as $\chi^{(3)}_{B}$. These pathways which involve absorption, emission, absorption and emission are described by another type of optical field correlation function $\langle a^\dag a a^\dag a \rangle$. The field factors for pathways (v-viii) are $\langle{a^\dag_2 a_2 a^\dag_1 a_1}\rangle$, $\langle{a^\dag_2 a_1 a^\dag_1 a_2}\rangle$, $\langle{a^\dag_1 a_1 a^\dag_2 a_2}\rangle$, $\langle{a^\dag_1 a_2 a^\dag_2 a_1}\rangle$, respectively.
\par
In case of classical optical modes \cite{marx2008nos} both types of field correlation functions can be factorized and all eight field factors become identical: $\langle a_1\rangle\langle a_2\rangle\langle a_1^\dag\rangle\langle a_2^\dag\rangle\sim |\mathcal{E}_1|^2|\mathcal{E}_2|^2$. Here $|\mathcal{E}_j|^2$ is proportional to the classical intensity of mode $j$. Thus, all pathways contribute equally to the conventional pump-probe signal:
\begin{equation}
\label{EQ:2}
S^{(C)}(\omega_1,\omega_2) \sim N|\mathcal{E}_1|^2|\mathcal{E}_2|^2
\Im [{\chi^{(3)}_{A}(\omega_1,\omega_2)+\chi^{(3)}_{B}(\omega_1,\omega_2)}]
\end{equation}
Here $\Im$ is the imaginary part; expressions of all eight pathways contributing to $\chi^{(3)}_{A}$ and $\chi^{(3)}_{B}$ are given in the Appendix \ref{apI}. The factor $N$ is generic for incoherent heterodyne detected processes \cite{marx2008nos}.
\par
We next turn to an experiment performed with broadband, high-flux photon pairs generated by the parametric down-conversion \cite{gerry2005iqo,glauber2007qto,mandel1995oca} of a pump beam with amplitude $\mathcal{E}_p$. A pair of photons is produced by illuminating a nonlinear crystal with the pump beam. The pair is then placed at the input ports of the Mach-Zehnder interferometer which regulates the degree of entanglement (See Fig.\ref{FIG:3} in the Appendix \ref{apII}). This configuration is known to exhibit strong, nonclassical fourth-order interference effects and has been used to explore the entangled-state properties of the parametric down converter \cite{PhysRevA.64.043802}.
\par
Choosing the phase shift introduced in one of the arms of the interferometer as $\phi=\pi$ (maximum entanglement) and using equations \eqref{chiALLLR} in Appendix \ref{apII}, the field factor for group A pathways assume the form \cite{PhysRevA.64.043802}:
\begin{equation}
\label{EQ:6}
\langle a^\dag a^\dag a a \rangle \sim |\mathcal{E}_p|^2+|\mathcal{E}_p|^4
\end{equation}
Whereas for group B we have:
\begin{equation}
\label{EQ:6B}
\langle a^\dag a^\dag a a \rangle \sim |\mathcal{E}_p|^4
\end{equation}
At low pump intensity $|\mathcal{E}_p|^4 \ll |\mathcal{E}_p|^2$  and the quadratic term dominates. In this limit the entanglement suppresses group B pathways and the entangled signal is given solely by group A:
\begin{equation}
\label{EQ:4}
S^{(E)}(\omega_1,\omega_2) \sim N |\mathcal{E}_p|^2
\Im \chi^{(3)}_{A}(\omega_1,\omega_2)
\end{equation}
The correlation functions \eqref{EQ:6} and \eqref{EQ:6B} can be controlled by the interferometer phase shift $\phi$. For example for $\phi=\pi/2$, both scale as $\sim |\mathcal{E}_p|^4$ and we recover the classical signal \eqref{EQ:2}
\par
The linear scaling of the signal with pump intensity $\sim |\mathcal{E}_p|^2$ has been predicted \cite{javanainen1990lid,PhysRevA.31.2409,dayan2007ttp} and verified experimentally \cite{lee2006epa,pe'er:073601} for various non-linear techniques and types of entanglements between photons. In the sum frequency generation technique only one pathway contributes to the signal and the matter enters through the second order susceptibility $\chi^{(2)}$, whether the modes are classical or entangled \footnote{The deviation from the $\chi^{(2)}$ description has been proposed by utilizing different propagations of the photons constituting the bi-photon. However, for thin nonlinear crystals, when one can achieve a substantial photon flux \cite{dayan:043602}, this effect is negligible.}. Hence, no new matter information is revealed by entangled pairs. The pump-probe technique proposed here, in contrast, shows clear pathway selectivity.
\begin{figure}
  \includegraphics[width=10cm]{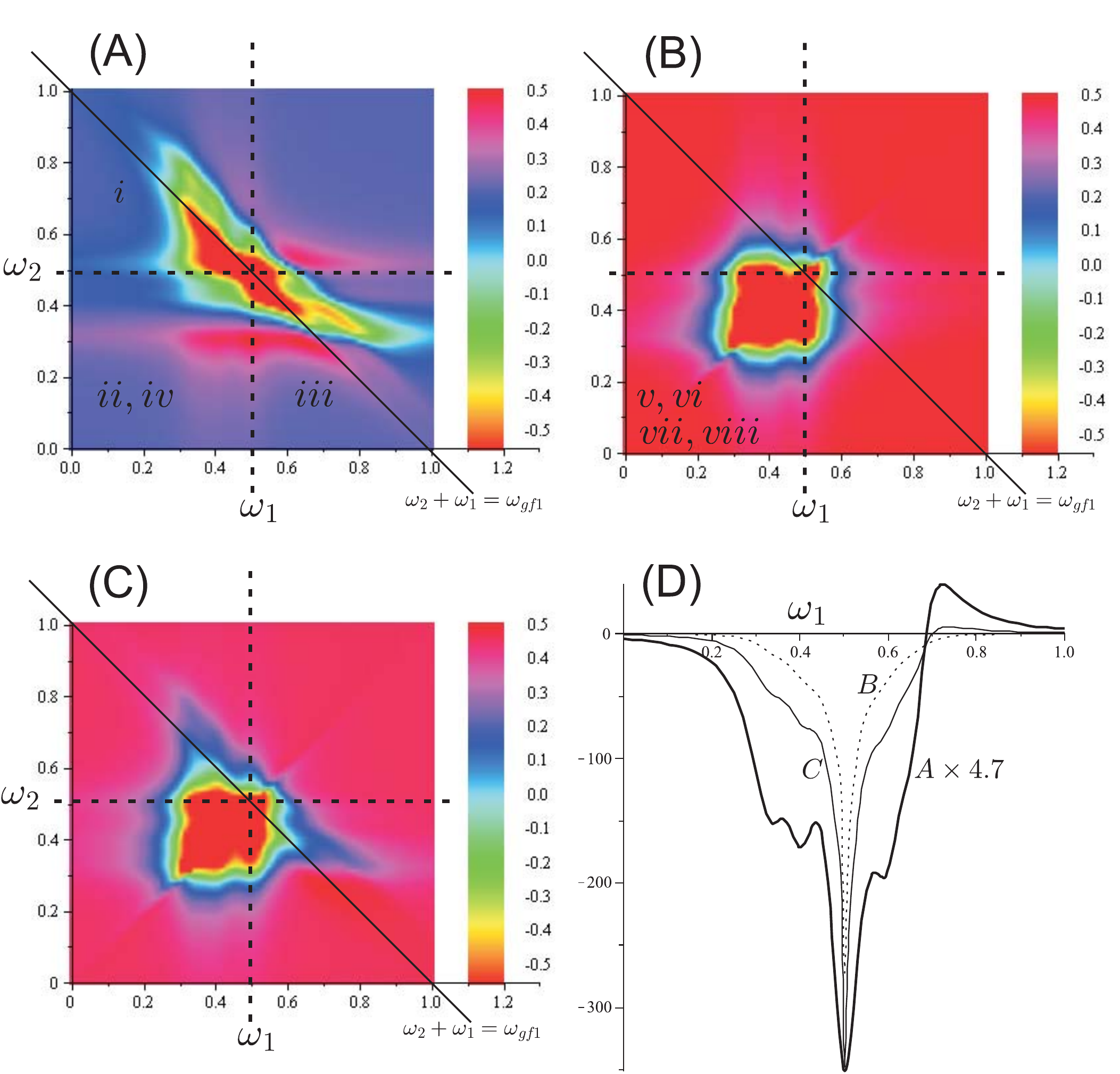}
  \caption{(Color on-line) (A) The entangled ($\phi=\pi$) signal $S^{(E)}(\omega_1,\omega_2)$ in eq.\eqref{EQ:4} which contains type A (TPA) pathways. (B) The signal from the SRS pathways of group B. (C) The classical pump-probe signal $S^{(C)}(\omega_1,\omega_2)$ ($\phi=\pi/2$) given by eq. \eqref{EQ:2}(sum of (A) and (B)). (D) Thick solid line - section of the entangled TPA $S^{(E)}(\omega_1,\omega_{gf1}-\omega_1)$; thin solid curve - the section of the classical pump-probe signal $S^{(C)}(\omega_1,\omega_{gf1}-\omega_1)$; dotted curve - group B (SRS) pathways contribution. These sections are marked by a diagonal line in panels (A)-(C).}\label{FIG:2}
\end{figure}
\par
Our three level ladder model (Fig.\ref{FIG:1}) has the following parameters: optical transition frequencies ($\omega_{eg}, \omega_{gf}$), transition dipole moments $\mu_{eg}, \mu_{ef}, \mu_{gf}$(as matrix elements of the dipole transition operator $V$) and the dephasing rate $\gamma$. In the following simulation we have used the following material parameters. The $|e\rangle$ states manifold is represented by three optical transitions $\omega_{eg}=\{10000,12000,15000\} \; cm^{-1}$. The $|f\rangle$ manifold has two two optical transitions $\omega_{fg}=\{ 30000, 33000\} \; cm^{-1}$.
The dephasing rate is $\gamma=1200 \; cm^{-1}$. The dipole moments of all the transitions are $\mu_{ge}=\mu_{ef}=\mu_{gf}=0.1$ (arbitrary units). For these parameters the group A and group B  signals overlap spectrally and may not be separated by classical fields.
\par
Simulated signals are shown in Fig.\ref{FIG:2}. In group A (Fig.\ref{FIG:2}(A)) the cross peaks due to pathway $i$ are given by the following  double resonance condition: $\omega_2\approx \omega_{eg}$ and $\omega_1+\omega_2\approx \omega_{fg}$ (upper left quadrant in Fig.\ref{FIG:2}(A)). For $iii$ they are given at $\omega_1 \approx \omega_{eg}$ and $\omega_1+\omega_2\approx \omega_{fg}$ (lower right quadrant in Fig.\ref{FIG:2}(A)). The cross-peaks from $ii$ and $iv$ pathways are given by triple resonance conditions. For $\omega_{eg} \neq \omega_{fg}$ they are much weaker (lower left quadrant in Fig.\ref{FIG:2}(A)). Group B pathways (Fig.\ref{FIG:2}(B)) show cross-peaks at $\omega_1\approx\omega_{eg}$ and $\omega_1 \approx\omega_{e'g}$. All of these pathways contribute in the lower left quadrant and interfere constructively. Since the signals of group A
and group B overlap spectrally (we assumed a large dephasing rate $\gamma/\omega_{eg}\approx 0.1$), the pathways may not be separated by classical optical fields, as is clear from the overall classical signal (their sum) shown in Fig.\ref{FIG:2}(C).
\par
The entangled photon signal shown in Fig.\ref{FIG:2}(A) can be used to unravel the TPA transitions $\omega_{ef}$ otherwise masked by the SRS transitions (Fig.\ref{FIG:2}(C)). To illustrate how this works, we display the diagonal ($\omega_2+\omega_1=\omega_{gf1} = 1$) section of the 2D spectrum (marked by diagonal solid line in panels A,B,C) in Fig.\ref{FIG:2}(D). For $\omega_1<0.5$ we see resonances at $\omega_1=\omega_{eg}$ (pathway $i$). For $\omega_1>0.5$ we see resonances at $\omega_{ef}$ given by pathway $iii$. At $\omega_1=0.5$ the resonance is enhanced by the
$ii$ and $iv$ pathways. The $\omega_{ef}$ resonances are very sensitive to the overlap between groups A and B on account of destructive interference between pathway $iii$ and pathways in group B. The resonances at $\omega_{eg}$ on the other hand are given by pathway $i$ and interfere constructively with group B. They gradually disappear as the dephasing rate is increased (not shown).
\par
In summary, under resonant conditions the matter does participate actively in the nonlinear processes and it becomes entangled with the
photons. For our parameter, in the conventional pump-probe experiment the TPA signal is masked by the SRS. Entangled photons eliminate the latter contributions at low pump intensity, and can resolve the $\omega_{ef}$ transitions.
\par
The fact that various matter paths are multiplied by different field paths creates an entanglement of matter and field paths. This paves the way for manipulating matter paths by controlling the field paths. The pump-probe experiment considered here singles out TPA pathways and eliminate SRS paths. Pathway selectivity can be used for studying complex material systems by revealing new matter information, otherwise erased by interference. Pathway manipulation with photon
entanglement is not limited to the TPA technique, but should be observed in other non-linear techniques such as difference frequency generation which involve four point optical field correlation function. Other types of entangled states may be used to achieve different kinds of selectivity. Sequences of short pulses of entangled photons may further be used to control the temporal material response. This could have numerous photonics and spectroscopic applications.

\begin{acknowledgments}
This work was supported by the National Institutes of Health Grant GM59230 and National Science Foundation Grant CHE-0745892.
\end{acknowledgments}
\appendix
\section{Material pathways in the pump-probe experiment. \label{apI}}
We introduce the auxiliary function:
\begin{equation}
I_{ab}(\omega)
=\frac{1}{\omega-\omega_{ab}+i \gamma}
\end{equation}
\par
The first two pathways of group A where all the interactions occur
with the ket make the following contribution:
\begin{gather}
\label{chiALLLL}
\chi^{(i)}_A(\omega_1,\omega_2)+\chi^{(ii)}_A(\omega_1,\omega_2)=\\
\notag
= \frac{-1}{3!\hbar^3} \sum \limits_{e} \sum \limits_{f} \mu_{ge'}\mu_{e'f} \mu_{fe}\mu_{eg}
[I_{eg}(\omega_2)I_{fg}(\omega_1+\omega_2)I_{e'g}(\omega_2)+
I_{eg}(\omega_1)I_{fg}(\omega_1+\omega_2)I_{e'g}(\omega_2)]
\end{gather}
The other two group A pathways where three interactions occur with
the ket and one with the bra give:
\begin{gather}
\label{chiALLLR}
\chi^{(iii)}_A(\omega_1,\omega_2)+\chi^{(iv)}_A(\omega_1,\omega_2)=\\
\notag
= \frac{1}{3!\hbar^3} \sum \limits_{e} \sum \limits_{f}\mu_{e'g}\mu_{fe'} \mu_{ef}\mu_{ge}
[I_{e'g}(\omega_2)I_{fg}(\omega_1+\omega_2)I^\star_{eg}(\omega_1)+
I_{e'g}(\omega_2)I_{fg}(\omega_1+\omega_2)I^\star_{eg}(\omega_1)]
\end{gather}
\par
The two group B pathways where all the interactions occur with the ket are:
\begin{gather}
\label{chiBLLLL}
\chi^{(v)}_B(\omega_1,\omega_2)+\chi^{(vi)}_B(\omega_1,\omega_2)=\\
\notag
\frac{-1}{3!\hbar^3}\sum \limits_{e} \sum \limits_{f} \mu_{ge'}\mu_{e'g} \mu_{ge}\mu_{eg}
[I_{eg}(\omega_1)I_{gg}(\omega_1-\omega_1)I_{e'g}(\omega_2)+
I_{eg}(\omega_2)I_{gg}(\omega_2-\omega_1)I_{e'g}(\omega_2)]
\end{gather}
The remaining two group B pathways where two interactions
occur with the ket and two with the bra are:
\begin{gather}
\label{chiBLLRR}
\chi^{(vii)}_B(\omega_1,\omega_2)+\chi^{(viii)}_B(\omega_1,\omega_2)=\\
\notag
= \frac{-1}{3!\hbar^3} \sum \limits_{e} \sum \limits_{f} \mu_{e'g}\mu_{ge'} \mu_{eg}\mu_{ge}
[I_{e'g}(\omega_2)I^\star_{gg}(\omega_2-\omega_2)I^\star_{eg}(\omega_1)+
I_{e'g}(\omega_1)I^\star_{gg}(\omega_1-\omega_2)I^\star_{eg}(\omega_1)]
\end{gather}
\par
Equations \eqref{chiALLLL}, \eqref{chiALLLR}, \eqref{chiBLLLL} and \eqref{chiBLLRR} were employed in simulating the classical and entangled signals:
\begin{gather}
\label{EQ:10}
S^{(C)}(\omega_1,\omega_2)=
\frac{N}{4\pi \hbar} |\mathcal{E}_1|^2|\mathcal{E}_2|^2
\Im [{\chi^{(3)}_{A}(\omega_1,\omega_2)+\chi^{(3)}_{B}(\omega_1,\omega_2)}]\\
\notag
S^{(E)}(\omega_1,\omega_2)=\frac{2N}{\pi \hbar} \left({\frac{2 \pi \hbar}{\Omega}}\right)^2\omega_1 \omega_2 \nu^2
\Im \chi^{(3)}_{A}(\omega_1,\omega_2)
\end{gather}

\section{Optical pathways and corresponding fourth order field correlation functions for classical and entangled photons. \label{apII}}
In a classical pump-probe experiment \cite{marx2008nos} the initial state of the optical field is a product of two coherent
states and the system initial condition is:
\begin{equation}
\label{EQ:1}
| t=-\infty \rangle = | g \rangle |\beta \rangle_1 |\beta \rangle_2
\end{equation}
\par
The electric field of a coherent mode $|\beta \rangle_\alpha$ is: $\mathcal{E}_\alpha=\beta_\alpha \sqrt{2 \pi \hbar \omega_\alpha/\Omega}$, where $\Omega$ is the mode quantization volume. In this case the optical correlation functions for all eight pathways scale as $\langle {a_1}\rangle\langle {a^\dag_1}\rangle\langle {a_2}\rangle\langle {a^\dag_2}\rangle \sim |\mathcal{E}_1|^2|\mathcal{E}_2|^2$. All pathways contribute equally to the classical pump-probe signal.
\par
\begin{figure}
  \includegraphics[width=8cm]{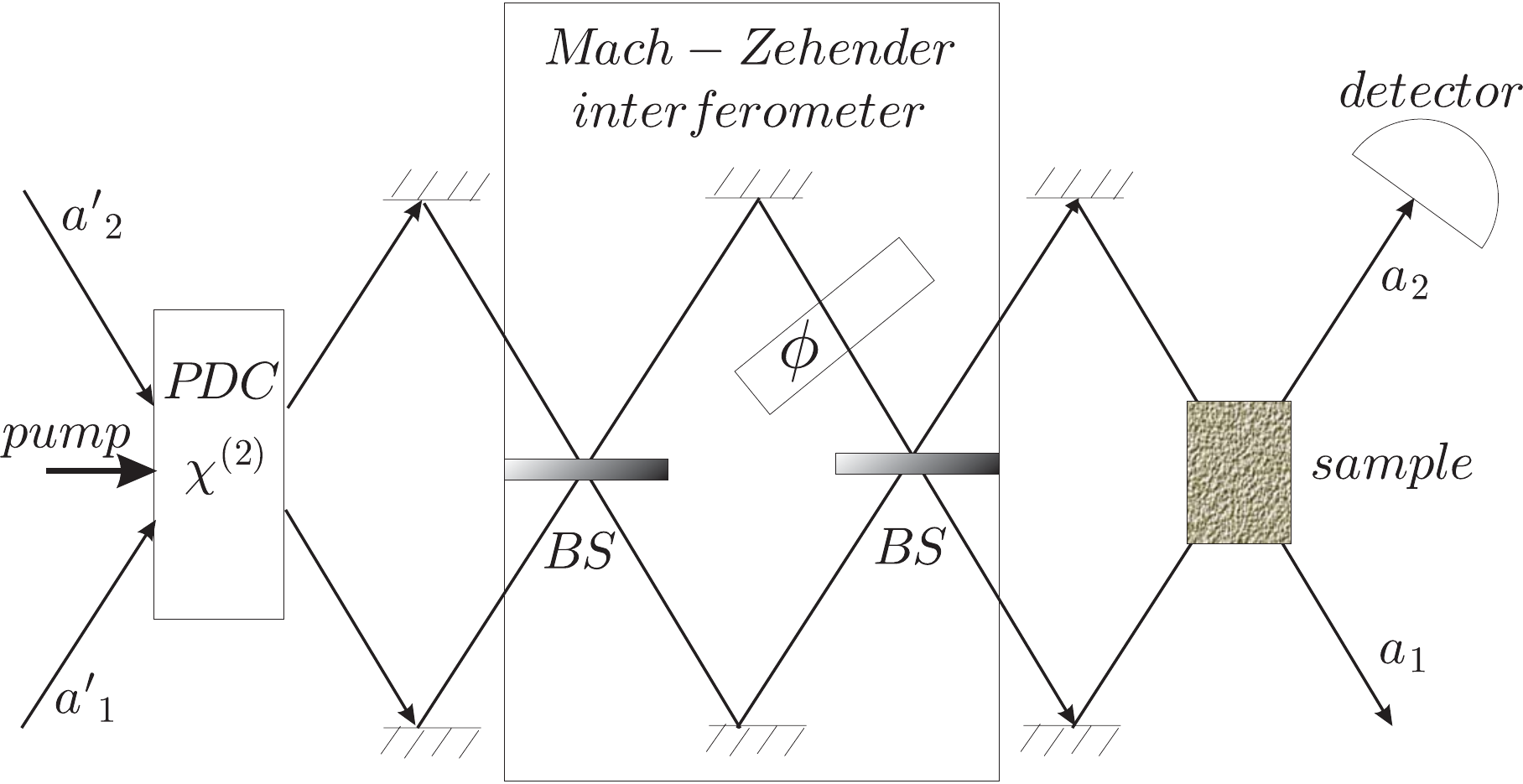}
  \caption{(Color on-line)Schematic of the pump-probe experiment with entangled photons. PDC is a non-linear $\chi^{(2)}$ crystal used to obtain entangled photon pairs from the classical pump beam by parametric down conversion. BS are the $50:50$ beam splitters. $\phi$ is a phase shift in one of the interferometer arms. The sample is a collection of $N$ three-level molecules. The detector measures the photon flux in the $\mathbf{k}_2$ mode. ${a'}_1,{a'}_2$ are annihilation operators for the original non-entangled (canonical) modes and $a_1,a_2$ represent the entangled modes.}\label{FIG:3}
\end{figure}

In an entangled-photon pump-probe experiment, the photon pairs are created by sending a single pump beam into a bi-refringent crystal. The two beams then pass through the Mach-Zehnder interferometer made of two beam splitters which mix the radiation modes (See Fig.\ref{FIG:3}).
The shift $\phi$ in one of the interferometer arms can further control the degree of entanglement. The state of the input field for the PDC + interferometer optical system is a product of two vacuum states $|0'\rangle_{1}|0'\rangle_{2}$. This optical system defines new modes $a_1,a_2$ which are related to the input (canonical) modes ${a'}_1,{a'}_2$ by a non-unitary transformation \cite{PhysRevA.64.043802}:
\begin{gather}
\label{transform}
a_1=\frac{1}{2}[{(1-e^{i\phi})(U{a'}_1+V{a'}_2^\dag)-i(1+e^{i\phi})(U{a'}_2+V{a'}_1^\dag)}]\\
\notag
a_2=\frac{1}{2}[{-i(1+e^{i\phi})(U{a'}_1+V{a'}_2^\dag)-(1-e^{i\phi})(U{a'}_2+V{a'}_1^\dag)}]
\end{gather}
Here $V=-i \sinh \nu, U=\cosh \nu$ with the parameter $\nu \sim \chi^{(2)} \mathcal{E}_p L$ determined by the crystal nonlinearity
$\chi^{(2)}$, the pump electric field $\mathcal{E}_p$ and the interaction path length $L$. 
\par
The pump-probe experiment is conducted with these transformed modes which can be distinguished by their wave vectors. The optical field correlation function corresponding to pathway $i$ can now be calculated and assumes the form:
\begin{gather*}
\langle 0',0'| a_2^\dag a_1^\dag a_1 a_2 |0',0'\rangle=\\
|V|^2[(\frac{1}{2}+\frac{1}{2}\cos 2 \phi)+|V|^2(\frac{3}{2}+\frac{1}{2}\cos 2 \phi)]
\end{gather*} All other Group A pathways give identical field factors:
$
\langle{a^\dag_2 a^\dag_1 a_1 a_2}\rangle=\langle{a^\dag_2 a^\dag_1 a_2 a_1}\rangle
=\langle{a^\dag_1 a^\dag_2 a_2 a_1}\rangle=\langle{a^\dag_1 a^\dag_2 a_1 a_2}\rangle
$.
\par
The sequential single photon absorption correlation function corresponding to pathway $v$ is:
\begin{gather*}
\langle 0',0'| a_2^\dag a_2 a^\dag_1 a_1 |0',0' \rangle=|V|^4(\frac{3}{2}+\frac{1}{2}\cos 2 \phi)
\end{gather*}
All other group B pathways optical correlation functions are the same:
$
\langle{a^\dag_2 a_2 a^\dag_1 a_1}\rangle=\langle{a^\dag_2 a_1 a^\dag_1 a_2}\rangle=
\langle{a^\dag_1 a_1 a^\dag_2 a_2}\rangle=\langle{a^\dag_1 a_2 a^\dag_2 a_1}\rangle
$.
\par
At low pump intensity $|\mathcal{E}_p|^2 \ll 1$ sequential single photon pathways of group B scale quadratically with pump intensity $|V|^4 \sim |\mathcal{E}_p|^4$. Their contribution to the signal may be neglected compared to bi-photon pathways which scale linearly with pump intensity $|V|^2(1+|V|^2)\sim |\mathcal{E}_p|^2$. The results of this appendix are used in equations \eqref{EQ:6} and \eqref{EQ:6B} of the main text.

\end{document}